\def\bea{\begin{eqnarray}}
\def\eea{\end{eqnarray}}

\def\beq{\begin{equation}}
\def\eeq{\end{equation}}
\def\ba{\beq\begin{array}{c}}
\def\ea{\end{array}\eeq}

  \newcommand{\bn}{\begin{enumerate}}
\newcommand{\en}{\end{enumerate}}
\newcommand{\bi}{\begin{itemize}}
\newcommand{\ei}{\end{itemize}}

\def\hat{\widehat}

\def\cech{${\rm C}^{\kern-6pt \vbox{\hbox{$\scriptscriptstyle\vee$}\kern2.5pt}}${\rm ech}}
\def\Cech{${\sl C}^{\kern-6pt \vbox{\hbox{$\scriptscriptstyle\vee$}\kern2.5pt}}${\sl ech}}

\def\d{{\delta}}

\def\CD{{\cal D}}

\def\CO{{\cal O}}

\def\CS{{\cal S}}

     \def\CO{{\cal O}}

\def\d{{\bf d}}

\def\inv{^{-1}}

 \def\frac#1#2{{\textstyle{#1\over#2}}}
\def\inv{^{\raise.15ex\hbox{${\scriptscriptstyle -}$}\kern-.05em 1}}

\def\({\left(}
\def\){\right)}
\def\<{\left\langle\,}
\def\>{\, \right\rangle}
\def\[{\left[}
\def\]{\right]}

\documentclass[12pt,a4paper]{article}
\input epsf
\usepackage{amsmath}
\usepackage{amssymb}
\usepackage{graphicx}
\usepackage{mathrsfs}
\usepackage{ulem}

\newcommand{\Tr}{{\rm Tr \,}}

\newcommand{\bpi}{ \mbox{\boldmath${\pi}$}}

\begin{document}
\title{Eigenvectors and scalar products for long range interacting spin chains II: the finite size effects}
\author{Didina Serban,\\ {\small \it  Institut de Physique Th\'eorique, DSM, CEA, URA2306 CNRS,} \\
 {\small \it Saclay, F-91191 Gif-sur-Yvette, France}}

\maketitle

\abstract{ In this note, we study the eigenvectors and the scalar products the integrable long-range deformation of the XXX spin chain 
defined in \cite{DSnote}. The model is solved exactly by algebraic Bethe ansatz, and it coincides in the bulk with the Inozemtsev spin chain. 
At the closing point it contains a defect  which effectively removes the wrapping interactions.
Here we concentrate on determining the defect term for the first non-trivial order in perturbation in the deformation parameter and how it affects the Bethe ansatz equations.
Our study is motivated by the relation with the dilatation operator of the ${\cal N}=4$ gauge theory in the $su(2)$ sector.
 } 
\bigskip

{\bf Introduction: Long range spin chains.}
In this paper we consider long-range integrable deformations of the XXX spin-1/2 spin chain. There are several methods to turn a nearest-neighbor spin chain into a long-range ones. One method, used in \cite{BBL1,BBL},
is to deform the conserved charges according to
\begin{equation}
\label{XXXdef}
\frac{d}{d\lambda}Q_r(\lambda)=i\left[ \chi(\lambda),Q_r(\lambda)\right]\;,
\end{equation}
where $\lambda$ the deformation parameter and  $\chi(\lambda)$ some deformation operator, which in the simplest case can be the boost operator of one of the higher conserved charges. The integrated version of the formula above gives
\begin{equation}
\label{Qdef}
Q_r(\lambda)=S (\lambda)Q_r(0)S^{-1} (\lambda)\;, \quad {\rm with} \quad \frac{dS}{d\lambda}\; S^{-1}=i\chi(\lambda)\;.
\end{equation}
This procedure can be realized on chains of infinite length, but on chains of finite size there are two complications. 
First, the definition of the operator $S (\lambda)$
might not be compatible with periodic boundary conditions, so the resulting chain will not be periodic.  Second, the transformation (\ref{Qdef}) cannot change the spectrum of a finite-dimesional system, so that the deformed chain will have the same spectrum as the undeformed one. On an infinite spin chain, we have 
in principle the possibility to use a singular transformation, which will thus change the spectrum. 
The generators of the symmetry algebra $J^a$ will transform in the same way as the charges,
\begin{equation}
\label{Jdef}
J^a(\lambda)=S (\lambda)J^a(0)S^{-1} (\lambda)\;.
\end{equation}
so the deformation (\ref{Qdef}) is a morphism of the symmetry algebra\footnote{We thank I. Kostov for this observation.}.
In particular, if the undeformed spin chain has Yangian invariance, this should be also the case for the deformed spin chain. The existence of an exact Yangian symmetry implies that there exists a monodromy matrix satisfying the Yang-Baxter equation with a rational $R$-matrix. In this case, we can use the algebraic Bethe ansatz method to diagonalize the transfer matrix and construct the eigenvectors and their scalar products.
An alternative way to proceed is to define the long range spin chains non-perturbatively  in terms of the deformation parameter, as compared to the perturbative definition implied by (\ref{XXXdef}). One example of chain defined non-perturbatively is the Inozemtsev model \cite {ino02}. It is known that this model, defined with periodic boundary conditions, is diagonalizable by (asymptotic) Bethe ansatz in the long chain limit. In the finite size limit the solution is more complicated, due to wrapping interactions. The absence of an exact solution in terms of Bethe ansatz is a sign that the Yangian symmetry algebra is not preserved by the finite size periodic chain.
In \cite{DSnote} we have defined a model which coincides in the bulk, in perturbation, with the Inozemtsev model but it differs from it by some defect term. 
The defect interaction is long range, with the range growing with the perturbation order, and effectively suppresses the wrapping interactions.
As a consequence,  the defect Inozemtsev spin chain  preserves exactly the Yangian algebra and  is diagonalizable by Bethe ansatz at any value of the deformation parameter and the length. 
The eigenstates and the scalar products are therefore computable by the usual algebraic Bethe ansatz procedure, as shown in  \cite{DSnote}.  
In this note, we work out the first non-trivial order in the defect term of the Hamiltonian and the corresponding Bethe ansatz equations up to next order
in the deformation parameter.
For long spin chains and low perturbation orders, the contribution of the defect is negligible and we obtain back the bulk Inozemtsev quantities. We compare with the results obtained from the boost deformations of the finite periodic chain.

{\bf The model.}
In \cite{DSnote} we have studied a long range Hamiltonian which can be constructed from the Dunkl operators \cite{Dunkl,Polychronakos:1992zk}
\begin{equation}
\label{DunklIno}
d_i^I=\sum_{j=i+1}^L\Theta_{ij}K_{ij}-\sum_{j=1}^{i-1}\Theta_{ji}K_{ij}=\sum_{j;j\neq i}^L\Theta_{ij}K_{ij}-\sum_{j=1}^{i-1}K_{ij}
\end{equation}
where $\Theta_{ij}=z_i/(z_i-z_j)$ and $z_j=e^{2j\kappa}$ with $\kappa$ a real number. The operators $K_{ij}$ permute the coordinates $K_{ij}z_j=z_i K_{ij}$.
Under the coordinate permutations, the Dunkl operators obey\footnote{In the defining relations of the degenerate affine Hecke algebra we use no index for the Dunkl operator, since they do not depend on the specific representation.}
 the relations of a degenerate affine Hecke algebra \cite{Cherednik}
\begin{eqnarray}
\label{Dunklperm}
&&\left[ d_i, d_j \right]=0\;,\quad 
K_{i,i+1}d_i-d_{i+1}K_{i,i+1}=1\;, \nonumber \\ \ 
&&[K_{i,i+1},d_k]=0\quad {\rm if}\ \ \  k\neq i,i+1 \;.
\end{eqnarray}
The model is defined via the monodromy matrix
\begin{equation}
\label{mmI}
T_a(u)\equiv \bpi (\hat T_a(u))\;, \qquad  \hat T_a(u)=\prod_{j=1}^L \(u-i/2-i\,d_j+iP_{ja}\)\;, 
\end{equation}
where the projection $\bpi$ transform coordinate permutations $K_{ij}$ into spin permutations $P_{ij}$ at the right of an expression,
\begin{equation}
\label{proj}
\bpi(\ldots K_{ij})=\bpi(\ldots)P_{ij}\;.
\end{equation}
Using the defining relations (\ref{Dunklperm}) of the degenerate affine Hecke algebra, one can prove that the Bernard-Gaudin-Haldane-Pasquier (BGHP)  \cite{BGHP} projection defined above is a morphism of the Yangian algebra
\begin{equation}
\label{morphism}
\bpi (\hat T_a(u)\hat T_{a'}(v))=\bpi(\hat T_a(u))\bpi(\hat T_{a'}(v))\;,
\end{equation}
and that $T_a(u)$ satisfies the Yang-Baxter equation with the rational $R$ matrix $R=u+iP$. 
The Hamiltonian associated to the model defined above is integrable, and when $L$ is large it coincides, in the bulk, with the Inozemtsev Hamiltonian \cite{ino02}
\begin{equation}
\label{Inoper}
H^I=4\sum_{i\neq j}\Theta_{ij}\Theta_{ji}(P_{ij}-1)=\sum_{i\neq j}\frac{1-P_{ij}}{\sinh^2\kappa(i-j)}\;.
\end{equation}
When $\kappa$ is large, the interaction falls off rapidly and only the nearest neighbors interact. The $\kappa\to \infty $ limit coincides thus with the Heisenberg model. We can use $\kappa$ as a tunable parameter to deform the XXX spin chain into a long range spin chain. This was done in \cite{SS} to reproduce the first three non-trivial orders in the perturbative expansion of the dilatation
of the ${\cal N}=4$ SYM theory in the $su(2)$ sector. The deformation parameter is given by the identification
\begin{equation}
\label{kappag}
t\equiv e^{-2\kappa}=g^2-3g^4+\CO(g^6)\;,
\end{equation}
where $16\pi^2g^2=\lambda$, the 't Hooft coupling constant of the gauge theory.
Since the sum in (\ref{DunklIno}) is finite and the coefficients $\Theta_{ij}$ are not periodic in $i\to i+L$, the Hamiltonian obtained from the above Dunkl operators is not translationally invariant. Instead, it can be seen as a closed chain with a defect around the sites $i=1$ and $i=L$. More general long range spin chain could be built in the same manner if we can find other representations of the Dunkl operators. In general, long range spin chains can be seen as reductions of 
short-range interaction systems by freezing some degrees of freedom. One particular example is the derivation of the BDS spin chain \cite{BDS} from the Hubbard model at half filling \cite{RSS}. In that example, one obtains a spin system with inhomogeneities $\theta_i=2g\sin q_i$ where $q_i$ are the momenta of the underlying fermions. The Dunkl operators can be thus seen as some dynamical impurities.
On the technical side, the interest in working with the operators (\ref{DunklIno}) is that they allow to fully exploit the algebraic Bethe ansatz formalism to construct the (eigen)vectors and their inner products.
The construction of the eigenvectors and of the conserved quantities for the long-range model (\ref{mmI}) is facilitated by the link  \cite{DSnote}  with the inhomogeneous XXX model with monodromy matrix
\begin{equation}
\label{mmXXX}
 T_0(u;\theta)=\prod_{j=1}^L \(u-\theta_j-i/2+iP_{ja}\)\;.
\end{equation}
The relation between the two objects is realized by the operator ${\mathscr{D}}_\theta$ defined by
\begin{eqnarray}
\label{thetamor}
T(u)&=&{\mathscr{D}}_\theta \;T_0(u;\theta) \Big\vert_{\theta=0}\\
&\equiv&\sum_n\sum_{j_1<\ldots<j_n}\sum_{k_1,\ldots,k_n} \frac{i^{k_1+\ldots k_n}}{k_1!\ldots k_n!}\; 
\;\partial_{j_1}^{\,k_1}\ldots  \partial_{j_n}^{\,k_n} \;T_0(u;\theta) \Big\vert_{\theta=0} \;\d_{j_1}^{\,k_1}\ldots \d_{j_n}^{\,k_n}
\nonumber
\end{eqnarray}
where we have used the notation\footnote{One should keep in mind that in general $\d_i\cdot \d_j\equiv \bpi(d_i) \bpi(d_j) \neq \d_i\d_j\equiv \bpi(d_i d_j) $.}
\begin{eqnarray}
 \d_{j_1}^{\,k_1}\ldots \d_{j_n}^{\,k_n}\equiv\bpi \(d_{j_1}^{\,k_1}\ldots d_{j_n}^{\,k_n}\)\;.
\end{eqnarray}
The morphism property of the BGHP projection (\ref{morphism}) gets transferred to the operator ${\mathscr{D}}_\theta$,
\begin{eqnarray}
\label{tBGHPmor}
T_a(u)T_{a'}(v)={\mathscr{D}}_\theta \;T_{0,a}(u;\theta) \Big\vert_{\theta=0} {\mathscr{D}}_\theta \;T_{0,a'}(u;\theta) \Big\vert_{\theta=0}={\mathscr{D}}_\theta \;[T_{0,a}(u;\theta)T_{0,a'}(v;\theta)] \Big\vert_{\theta=0}\;.
\end{eqnarray}
Upon acting on the pseudo-vacuum state $|\Omega\rangle=|\uparrow\uparrow\ldots\uparrow\rangle$, the BGHP morphism becomes a differential operator, close to the theta quasi-morphism\footnote{We call the action of $\CD_\theta$ a quasi-morphism because it contains cross terms which spoil the morphism property.} defined in \cite{GV,GV2},
\begin{eqnarray}
\label{vacvac}
{\mathscr{D}}_\theta \;[\ldots](u;\theta) \Big\vert_{\theta=0}|\Omega\rangle={\cal{D}}_\theta \;[\ldots](u;\theta) \Big\vert_{\theta=0}|\Omega\rangle \;,
\end{eqnarray}
where the dots stand for any product of elements of the monodromy matrix, and
\begin{equation}
\label{thetaop}
\CD_\theta=\sum_n\sum_{j_1<\ldots<j_n}\sum_{k_1,\ldots,k_n} \frac{i^{k_1+\ldots k_n}}{k_1!\ldots k_n!}\; C^{k_1,\ldots,k_n}_{j_1,\ldots,j_n}
\;\partial_{j_1}^{\,k_1}\ldots  \partial_{j_n}^{\,k_n} \;,
\end{equation}
\begin{equation}
C^{k_1,\ldots,k_n}_{j_1,\ldots,j_n}\,|\Omega\rangle\equiv \d_{j_1}^{\,k_1}\ldots \d_{j_n}^{\,k_n}\,|\Omega\rangle\;.
\end{equation}
The coefficients $C^{k_1,\ldots,k_n}_{j_1,\ldots,j_n}$ can be computed in principle from the knowledge of the Dunkl operators. In \cite{DSnote},
we have computed their values in the bulk up to order $g^4$ in the perturbative expansion (\ref{kappag}), which are translationally invariant. If we take into account the effect of the non-periodicity of the definition (\ref{DunklIno}), we have
\begin{equation}
\CD^I_\theta=1+ig^2(\partial_{L}-\partial_1)+\frac{g^2}{2} \sum_{i=1}^{L-1}(\partial_{i+1}-\partial_i)^2+\CO(g^4)\;.
\end{equation}
 From the above identification we deduced that the (eigen) vectors and their scalar products for the long range model can be straightforwardly computed from the corresponding quantities in the inhomogeneous model. 
Denoting as usual with $A(u),\ B(u),\ C(u),\ D(u)$  the elements of the matrix $T(u)$, the eigenvectors of the transfer matrix, $\Tr T(u)=A(u)+D(u)$, 
can be constructed as
\begin{equation}
\label{thetavec}
|\{u\}\rangle _g\equiv B(u_1)\ldots B(u_M)|\Omega\rangle=\CD_\theta |\{u;\theta\}\rangle_{\theta=0} \;,
\end{equation}
where $|\{u;\theta\}\rangle=B_0(u_1;\theta)\ldots B_0(u_M;\theta)|\Omega\rangle$. Since $T(u)$ obeys the Yang-Baxter equation with the rational $R$
matrix, $R(u)=u+iP$, the algebra of the matrix elements is the same as for the usual XXX model,
\begin{eqnarray}
A(v)B(u)=\frac{u-v+i}{u-v} B(u) A(v)-\frac{i}{u-v} B(v) A(u)\;,\\
D(v)B(u)=\frac{u-v-i}{u-v} B(u) D(v)+\frac{i}{u-v} B(v) D(u)\;.
\end{eqnarray}
Therefore, the vectors $|\{u\}\rangle _g$ are eigenvectors of the transfer matrix $\Tr T(u)$ with eigenvalue
 \begin{eqnarray}
t(u)=a(u)\frac{Q(u-i)}{Q(u)}+d(u)\frac{Q(u+i)}{Q(u)}\;, \quad {\rm with} \quad Q(u)=\prod_{i=1}^M(u-u_i)\
\end{eqnarray}
provided that the rapidities $\{u\}$ are satisfying the Bethe ansatz equations
\begin{equation}
\frac{a(u_j)}{d(u_j)}=\prod_{k\neq j}^M\frac{u_j-u_k+i}{u_j-u_k-i}\;.
\end{equation}
Above, $a(u)$ and  $d(u)$ the are eigenvalues of $A(u)$ and $D(u)$ on the pseudo vacuum 
\begin{eqnarray}
&&a(u)=f(u+i/2)^L\;, \qquad d(u)=f(u-i/2)^L\;,\\
&& f(u)^L|\Omega\rangle=\CD_\theta \prod_{j=1}^L (u-\theta_j)\,|\Omega\rangle|_{\theta=0}=\bpi\left[ \prod_{j=1}^L (u-id_j)\right]|\Omega\rangle\;.
\end{eqnarray}
To determine the function $f(u)$ we only need the value of the projection of the symmetric sums,
 \begin{eqnarray}
&\sum_{k=1}^L\d_k^n \, |\Omega\rangle=(-i)^n L C_n \,|\Omega\rangle&
\end{eqnarray}
with $C_n$ determining the expansion  
\begin{eqnarray}
\frac{d }{du}\ln f(u)=\frac{1}{u}\sum_{n\geq 0}\frac{C_n}{u^n}\;.
\end{eqnarray}
By explicit computation using the expression (\ref{DunklIno}) and the perturbative expansion (\ref{kappag}) we get 
$C_{2n+1}=0$ and $C_0=1$, $C_2=2g^2(1-1/L)-2g^4/L+\CO(g^6)$ and $C_4=6g^4(1-5/3L)+\CO(g^6)$. The  
$L^0$ part gives the bulk contribution, which was already evaluated in \cite{DSnote}. 
Denoting with $x(u)$ the bulk part, which up to order $g^4$ is the same for Inozemtsev and BDS spin chain, we find that 
\begin{eqnarray}
f(u)^L=x(u)^L\(1+\frac{g^2+g^4}{u^2}+\frac{3g^4}{u^4}+\CO(g^6)\)\;.
\end{eqnarray}
It is remarkable that the correction does not depend on $L$. Up to the specified order, the Bethe ansatz equation can be written as
\begin{eqnarray}
\label{impBAE}
\(\frac{x_j^+}{x_j^-}\)^{L}=e^{i\phi_0(u_j)}\prod_{j\neq k}\frac{u_j-u_k+i}{u_j-u_k-i}\;,
\end{eqnarray}
with the phase shift $\phi_0(u)$ given by
\begin{eqnarray}
\label{phshdef}
\phi_0(u)=i(g^2+g^4)\(\frac{1}{(u^+)^2}-\frac{1}{(u^-)^2}\)+\frac{5g^4i}{2}\(\frac{1}{(u^+)^4}-\frac{1}{(u^-)^4}\)+\CO(g^6)\;,
\end{eqnarray}
where we used the notation $u^\pm=u\pm i/2$ and $x_j^\pm=x(u_j^\pm)$. The phase shift $\phi_0(u)$ in the Bethe ansatz equations  (\ref{impBAE}) can be interpreted as a scattering on a defect situated near the closing point of the chain, $k=L$.  This interpretation will become clearer when we shall derive the explicit form of the Hamiltonian, see below. A similar phase shift appeared in the treatment of long-range deformations of  boundary spin chains, \cite{Florian}
and it can be traced back to the odd conserved charges of the chain \cite{BBL1}.
Since the strength of the defect does not depend on $L$ (at least to the given order), on large chains the effect of the defect will be subleading, and at the leading order in $L$ we retrieve the bulk  Bethe ansatz equations.  Let us note that equations (\ref{impBAE}) are exact for any $L$, provided that we have determined $\phi_0$. For the Inozemtsev model, where we know the exact expression of the Dunkl operator, this can be done in principle, but we are not doing it here beyond three loop. 

Once the algebra of the matrix elements of the monodromy matrix is known, the scalar products of an eigenstate with an arbitrary off-shell state ({\it i.e.} with rapidities not satisfying Bethe ansatz equations)  can be computed  
by the  Slavnov  method \cite{Slavnov}, and it reduces to the determinant 
\begin{eqnarray}
\nonumber
_g\langle \{v\}|\{u\}\rangle_g
=\CS_{\{u\},\{v\}}=\prod_{j=1}^M a(v_j)\,d(u_j)\,\frac{\det_{jk}\Omega(u_j,v_k)}{\det_{jk}\frac{1}{u_j-v_k+i}}
\end{eqnarray} 
with
\begin{eqnarray}
&&\Omega(u,v)=t(u-v)-e^{2ip_u(v)}t(v-u)\;,\\
&& t(u)\equiv\frac{i}{u(u+i)}\;,  \qquad
\label{qmBDS} \nonumber
e^{2ip_{\bf u}(u)}\equiv\left[\frac{f(u^-)}{f(u^+)}\right]^L\prod_{j}^M\frac{u-u_j+i}{u-u_j-i}\;.
\end{eqnarray}
When $\{u\}=\{v\}$ this formula gives  the norm of the states in the algebraic Bethe ansatz normalization.

{\bf The conserved quantities.}
On an infinite lattice, the Inozemtsev Hamiltonian (\ref{Inoper}) can be built from the quantum determinant of the monodromy matrix, as explained in \cite{HT}.
On a finite size lattice  the Hamiltonian does not commute anymore with the Yangian, so 
it should not be obtained from the quantum determinant, but rather from the transfer matrix, by taking higher derivatives\footnote{A similar strategy was taken in \cite{GV2}.}  of $\ln T(u)$ at $u=i/2$. When $d_i=0$ we have the usual XXX conserved Hamiltonians
 \begin{eqnarray}
H^0_n=\frac{i^{n-1}}{(n-2)!}\; \partial^{n-1}_u \ln T_0(u)|_{u=i/2}\;. 
\end{eqnarray}
or, explicitly, for the first few charges
 \begin{eqnarray}
 H_1^0=\ln U_0\;, \qquad U_0=\Tr_a P_{a1}\ldots P_{aL}=P_{L,L-1} \ldots P_{12}
 \end{eqnarray}
  \begin{eqnarray}
H^0_2=U_0^{-1} \sum_k  \Tr_a P_{a1}\ldots \check P_{ak}\ldots P_{aL}=\sum_j P_{kk+1}\;, 
\end{eqnarray}
where the check means the corresponding factor is absent.
The next charges are given by the higher order derivatives
\begin{eqnarray}
H^0_3&=&\sum_{k=1}^L [P_{kk+1},P_{k+1,k+2}]-L\;,\\
H^0_4&=&\sum_{k=1}^{L}[P_{kk+1},[P_{k+1,k+2},P_{k+2,k+3}]]-\sum_{i=1}^{L}P_{k,k+2}+2\sum_{i=1}^LP_{k,k+1}\;. \nonumber
\end{eqnarray}
The eigenvalues of the transfer matrix are given by
\begin{eqnarray}
t_0(u)=a_0(u)\frac{Q(u-i)}{Q(u)}+d_0(u)\frac{Q(u+i)}{Q(u)}\;,
\end{eqnarray}
\begin{eqnarray}
{\rm with}\quad a_0(u)=(u+i/2)^L\;, \quad d_0(u)=(u-i/2)^L\;,
\end{eqnarray}
so that the eigenvalues of the $n$th Hamiltonian (except for $n=1$) are given by
\begin{eqnarray}
\label{ezerou}
E_n^0=(-1)^nL-\sum_{j=1}^M\left[\(\frac{i}{u_j^+}\)^{n-1}-\(\frac{i}{u_j^-}\)^{n-1}\right]
\end{eqnarray}
where we used the notation $u^\pm=u\pm i/2$.
Now we will turn to computing the conserved quantities of the deformed model (\ref{mmI}). When the $d_j$'s are present, we have
 \begin{eqnarray}
 U=\bpi\( \Tr_a \;(P_{a1}-d_1)\ldots (P_{aL}-d_L)\)\;,
 \end{eqnarray}
  \begin{eqnarray}
H_2=i\;\partial_u \ln T(u)\Big\vert_{u=i/2}
=\bpi \(U^{-1}\sum_j  \Tr_a \;(P_{a1}-d_1)\ldots \check{(P_{aj}-d_j)}\ldots  (P_{aL}-d_L)\)\;.\nonumber
\end{eqnarray}
Computing these operators by evaluating the projections explicitly might be involved . We are interested here in understanding 
the structure of the deformation in the perturbative parameter $g^2$ from  the equation (\ref{kappag}).
Expanding to the second order in the Dunkl operators we obtain an expression which is similar to the Hamiltonian with impurities \cite{BBL}, 
since before projection the Dunkls behave as c-numbers.
\begin{eqnarray}
\label{lntrtbis}
&&\qquad U=U_0\(1-\sum_k P_{kk+1}\d_k+\sum_{k<j} P_{kk+1}P_{jj+1}\d_k\d_j+\CO(\d^3)\)\;,\\ \nonumber
 &&\qquad H_2 
=\sum_k P_{kk+1}+\sum_k \d_k-\sum_{k} [P_{k-1,k}, P_{kk+1}]\,\d_k
 \\
\!\!\!\!\!\!\!\!\!\!\!\!&&+ \sum_{k}(P_{kk+1}+P_{k,k-1}-P_{kk+2})\,\d_k^2+\sum_{k}[[P_{k-1k},P_{kk+1}] , P_{k+1k+2}]\,\d_k\d_{k+1}+\CO(\d^3)  \nonumber
\end{eqnarray}
The result of the projection of the Dunkl operators can be computed by hand for the $g^2$ terms, and using Mathematica for the $g^4$ terms. Up to terms of order $g^4$  in the bulk we have
 \begin{eqnarray}
 \d_k=g^2(P_{kk-1}-P_{kk+1}), \quad \d_k^2=-2g^2,\quad \d_k\d_{k+1}=g^2
 \end{eqnarray}
and at the boundary  we have
 \begin{eqnarray}
 \d_1=-g^2P_{12}=\d_1^{\rm bulk}-g^2P_{L1},&\quad& \d_L=g^2P_{L-1,L}=\d_L^{\rm bulk}+g^2P_{L1}, \nonumber\\
  \d_1^2=\d_L^2=-g^2=\d_k^{2,{\rm bulk}}+g^2,&\quad& \d_L\d_1=0=(\d_L\d_{1})^{\rm bulk}-g^2\;.
  \end{eqnarray}
Replacing these values in (\ref{lntrtbis})  we get
\begin{eqnarray}
\nonumber
U&=&U_0\(1+g^2(L-1)+g^2\sum_k [P_{kk-1},P_{k,k+1}]-g^2[P_{L1},P_{12}]\)+\CO(g^4)\\
H_2  \nonumber
&=&H_2^0+g^2\sum_{k=1}^{L}[P_{kk-1},[P_{kk+1},P_{k+1,k+2}]]-2g^2\sum_{k=1}^{L}P_{kk+1}\\
&+&2g^2P_{L1}-g^2[P_{L-1,L}[P_{L1},P_{12}]]+\CO(g^4)\;.
\end{eqnarray}
If we now combine with  higher XXX Hamiltonians, necessary to get rid of the multispin bulk interaction, we get
\begin{eqnarray}
\label{shiftdef}
\ln U_{\rm def}&=&\ln U -g^2 H_{3}^0-(2L-1)g^2=\ln U_0-g^2[P_{L1},P_{12}]+\CO(g^4)\;,\\
H_{2, \rm def}&=&H_2-g^2H_4^0= \nonumber
(1-4g^2)\sum_{k=1}^LP_{kk+1}+g^2\sum_{k=1}^{L}P_{kk+2}\\
\label{hamdef}
&+&2g^2P_{L1}-g^2[P_{L-1,L}[P_{L1},P_{12}]]+\CO(g^4)\;.
\end{eqnarray}
The operator (\ref{shiftdef})  is the  momentum operator, while the Hamiltonian  (\ref{hamdef}) coincides with the bulk two-loop dilatation operator plus a  defect term.
The same bulk expressions can be obtained by deformation of the spin chain using the procedure from equation (\ref{XXXdef}),
\begin{eqnarray}
\chi(g^2)=B[H_3^0]+\CO(g^2)=i\sum_{k=0}^{L-1} k[ P_{k,k+1},P_{k+1,k+2}]+\CO(g^2)\;,
 \end{eqnarray}
 where $B[H_3^0]$ is called the boost operator of $H_3^0$.
As it is defined above, the boost operator also generates a defect term, but which is $L$ times the defect term in (\ref{shiftdef}) and (\ref{hamdef}). This can be understood from the fact that  the deformation (\ref{XXXdef}) is in fact a similarity transformation and it should not change the spectrum/Bethe equations on a finite chain.
Let us now characterize the spectrum of the conserved quantities of the defect Hamiltonians determined above. For the deformations we consider, $d(i/2)=x(0)^L$, so the conserved quantities, for long spin chains,
have the same functional form in terms of rapidities as in formula (\ref{ezerou}) of the XXX spin chain, up to some higher order corrections that we can neglect here.\footnote{In the BDS spin chain the corrections are of the order $g^L$.}.
This does not mean that the spectrum of the conserved charges is the same; the quantization conditions implied by the Bethe equations are different.
Moreover, we have redefined the conserved quantities in (\ref{shiftdef}) and (\ref{hamdef}) by taking linear combinations, so 
we get
\begin{eqnarray}
\label{edef}
E_{\rm def}=E_2-g^2E_4+\CO(g^4)&=&C-i\sum_{j=1}^M\(\frac{1}{u^+_j}+\frac{g^2}{(u_j^+)^3}-\frac{1}{u_j^-}-\frac{g^2}{(u_j^-)^3}\)+\CO(g^4)\nonumber \\
&= & C-\sum_{j=1}^M\(\frac{i}{x(u_j^+)}-\frac{i}{x(u_j^-)}\)+\CO(g^4)
\end{eqnarray}
with $C=iL(f(i)+2g^2f''(i))$.  In this way we obtain a perturbative expression which matches the charges for the BDS spin chain \cite{BDS}. 
From (\ref{shiftdef}) and the properties of the shift operator $U_0$ we conclude that 
\begin{eqnarray} (U_{\rm def})^L=1-g^2(H_3+L)+\CO(g^4)\;.
\end{eqnarray}
Let us now look at the one magnon states with wave function  given by 
\begin{eqnarray}
\Psi_p=\sum_{k=1}^L (e^{ipk}+g^2c_k)|k\rangle\;.
\end{eqnarray}
where we have denoted with $|k\rangle =\sigma_k^-|\Omega\rangle$. The coefficients  $c_k$ are zero in the bulk. Their value near the defect can be computed in the simplest way by imposing that $\Psi_p$ is an eigenvector of $U_{\rm def}$ with eigenvalue $e^{ip}$. We obtain
\begin{eqnarray}
c_L=e^{2ip_0}-e^{ip_0}\;, \quad c_1=e^{-ip_0}-1\;.
\end{eqnarray}
From the previous formula, we obtain, at  two loops,
\begin{eqnarray}
pL=2\pi n +ig^2\(\frac{1}{(u^+)^2}-\frac{1}{(u^-)^2}\)+\CO(g^4)
\end{eqnarray} 
so that the quantization condition for the momentum $p$ is
\begin{eqnarray}
&&pL-16g^2\cos p/2 \sin^3 p/2=2\pi n\;,\\ {\rm or } \quad &&p=p_0+\frac{16g^2}{L}\cos p_0/2 \sin^3 p_0/2\;, \nonumber \quad
{\rm with} \quad p_0=\frac{2\pi n}{L}\;.
\end{eqnarray}
At the next loop order, taking into account the expression (\ref{phshdef}) of the defect phase shift one has
\begin{equation}
p=p_0 +\frac{16g^2}{L}\sin^3\frac{ p_0}{2}\cos \frac{ p_0}{2}\left[1\!-\!g^2 \!\(\!
  4\!-\!10 \cos  p_0\! +\!5\cos  2p_0\!+\!\frac{4}{L}  (\cos  2p_0\!-\!\cos  p_0 )\!\)\!\right]. \nonumber
\end{equation}
The  vector $\Psi_p$ diagonalizes also the defect Hamiltonian (\ref{hamdef}) with eigenvalue given by (\ref{edef}). Let us emphasize that, although the
functional form of the eigenvalues (\ref{edef}) are the same as in the translational invariant case, the actual eigenvalues are modified due to the change
of the Bethe equation (quantization condition in the case of the one-magnon states),
\begin{eqnarray}
E_{2,{\rm def}}( p)&=&- 4\sin^2 p/2+16g^2\sin^4 p/2 +\CO(g^4)\\
&=&- 4\sin^2 p_0/2+16g^2\sin^4 p_0/2-\frac{64g^2}{L}\cos^2 p_0/2\sin^4p_0/2 +\CO(g^4)\;.
\nonumber
\end{eqnarray}
We have checked that the two-loop eigenfunctions can be obtained alternatively by applying the operator $\CD_\theta$ to the eigenvectors of the inhomogeneous XXX model, {\it cf.}
equation (\ref{thetavec}).  

{\bf Conclusion} We have studied the conserved quantities of a long-range deformation of the XXX model that is defined globally, {\it i.e} not only perturbatively.
This model coincides in the bulk with the Inozemtsev model \cite{ino02}, which in turns  coincides, up to three-loop order in the deformation parameter, with the BDS model \cite {BDS} and with the dilatation operator on the ${\cal N}=4$ SYM theory. The difference with the the Inozemtsev model is given by a defect term situated at the closing point of the chain. The exact spectrum and eigenvectors of the resulting Hamiltonian can be built by algebraic Bethe ansatz, for any length of the spin chain. On a chain of length $L$ they differ from the corresponding quantities of the periodic chain by terms of order $1/L$. This justifies their use in \cite{DSnote} to extend the semiclassical results of \cite{Ivan} to two-loop three-point correlation functions in the $su(2)$ sector of  ${\cal N}=4$ SYM \cite{Okuyama:2004bd,RV,EGSV}.

{\bf Acknowledgements.} We would like to thank N. Gromov, R. Janik, I. Kostov and F. Loebbert for  fruitful exchange and discussions, and the CQUEST Seoul, where part of this work was done,  for warm hospitality.
The research leading to these results has received funding from the  
[European Union] Seventh Framework Programme [FP7-People-2010-IRSES]  
under grant agreement no. 269217 and the Sakura exchange program PHC 27588UA of the French Ministry of Foreign Affairs.

\end{document}